# Correlations in mesoscopic magnetic systems

F. Gulminelli[1] and Ph. Chomaz[2]

LPC Caen, F-14050 CAEN Cédex FRANCE
GANIL, BP5027, F-14076 CAEN Cédex5 FRANCE

**Subject** :
The purpose of this proposal is to study the ferro/para phase transition in a mesoscopic Ising-like lattice and in particular demonstrate the existence of a negative magnetic susceptibility in the fixed magnetization ensemble.

**Experimental observable** :
To this aim we will use the correlation

$$< S(r)\, S(r+\Delta r) > \; = \; < M^2 >/N^2$$

where N is the total number of spins for a single cluster, M the total magnetization of the cluster, and the equality holds if we choose

$$r_0 < \Delta r < R$$

where $r_0$ is the linear size of a spin site and R is the linear size of a cluster.

**Introduction :**
Theory predicts that in the phase transition region for a finite lattice the equation of state relating the field to the magnetization presents a backbending, i.e. a negative susceptibility.

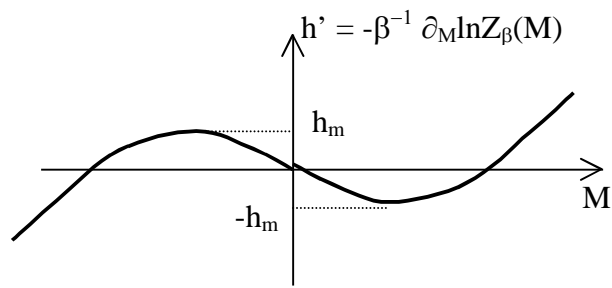

To directly measure this backbending one should be able to produce a statistical ensemble where the magnetization of each cluster of spins is exactly the same with no fluctuations (fixed magnetization ensemble). This is impossible to realize, however theory assures that a backbending is equivalent to a bimodal magnetization distribution : a non zero interval of magnetic field ($-h_m < h < h_m$) exists such that the magnetization distribution for a statistical ensemble subject to an external field is represented by two magnetization peaks separated by a barrier of very low probability.

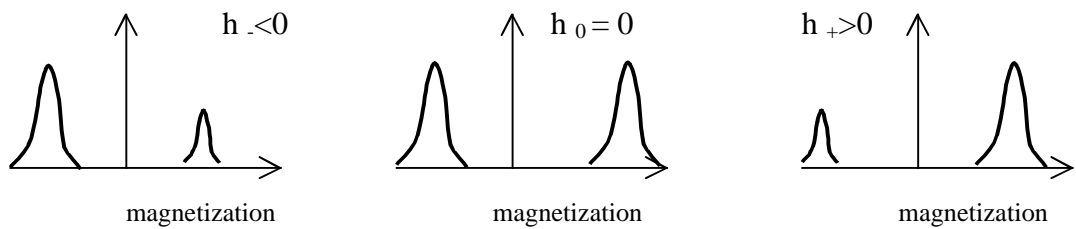

Because of the barrier between the two magnetization peaks, a short time experiment can explore the most probable solution as a metastable state, while a long time experiment will give rise to the complete equilibrium distribution.

The transition from the metastable to the stable equilibrium should then correspond to a strong increase of the variance of the magnetization distribution.

Since the magnetization variance is a measure of the susceptibility of the system subject to a field

$$\beta \sigma^2 = \beta (<M^2>-<M>^2) = \chi$$

the abnormally large fluctuation can be quantitatively associated to the anomalous susceptibility in the fixed magnetization ensemble.

## Experiment :

Let us start with an equilibrated sample at a magnetic field out of the phase transition $h>h_m$. This situation corresponds to a normal (monomodal) magnetization distribution with a mean value $M_h >0$ and a variance $\sigma_h$.

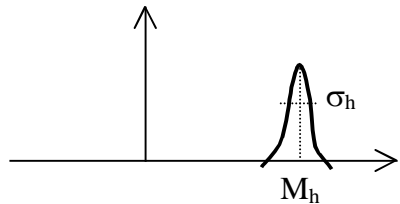

At $t = 0$ the magnetic field suddenly drops to a value $h_+<h_m$ inside the transition region. The magnetization distribution should first reach the metastable monomodal solution with mean value $M_+<M_h$ and variance $\sigma_+$, within a characteristic time $\tau$,

$$M-M_+=(M_h-M_+) e^{-t/\tau}$$

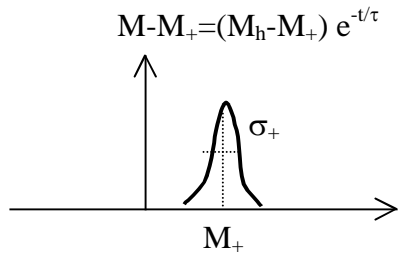

Over a longer time scale we should be able to observe the decay to the genuine equilibrium bimodal distribution with mean value $M_\infty$ and variance $\sigma_\infty$,

$$M-M_\infty=(M_+-M_\infty) e^{-t/\tau'} \quad \tau' >> \tau$$

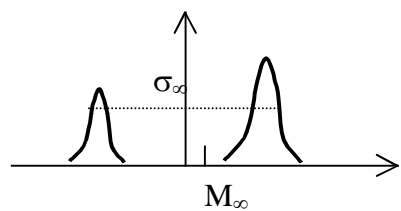

Here is a schematic representation of the experiment.

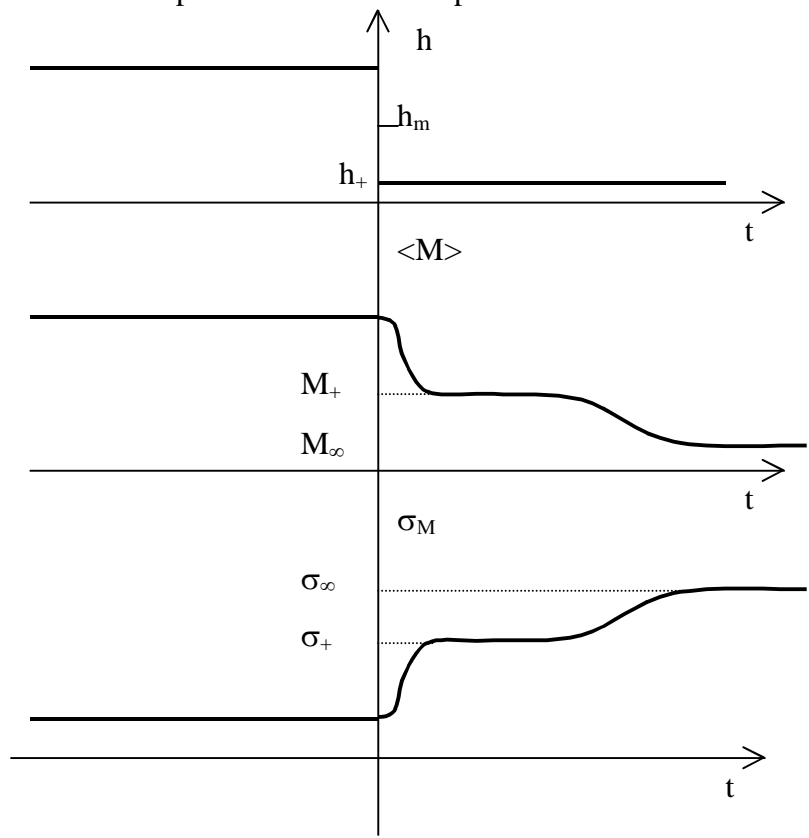

## Physical results :
The experiment can be repeated for different values of the asymptotic field $h_+$ , starting either from a positive or from a negative initial field.
The results should look like that :

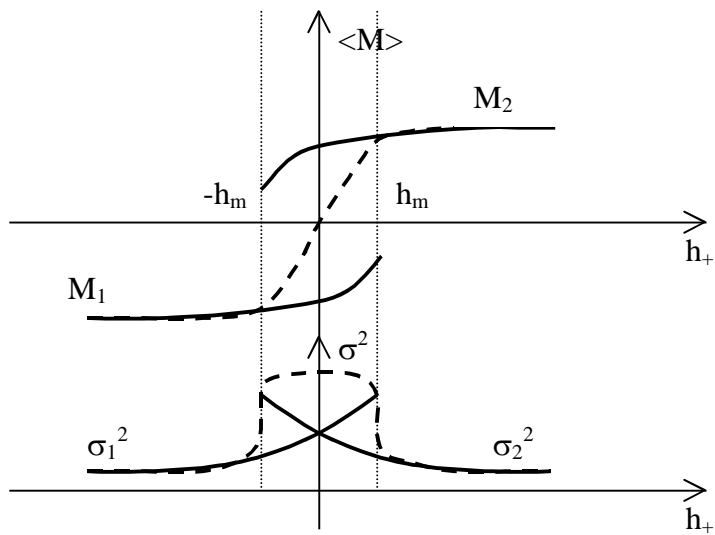

Where :
- The full lines ($M_1$, $M_2$, $\sigma_1^2$, $\sigma_2^2$) represent the metastable behaviors, the dashed line (<M>, $\sigma^2$) the final equilibrium ones.
- The left (right) metastable branch is obtained by starting from a negative (positive) initial field.

The measurement of the mean values allows to determine the equilibrium and metastable susceptibility under a magnetic field

$$\chi(h) = \partial \langle M \rangle / \partial h \ ; \ \chi_1(h) = \partial M_1 / \partial h \qquad (1)$$

that should agree with the measurement of the variances

$$\chi(h) = \beta \sigma^2(h) \ ; \ \chi_1(h) = \beta \sigma_1^2(h) \qquad (2)$$

where $\beta$ is the inverse of the temperature.

The measurement of the variance also allows to obtain the width of the backbending region in the constant magnetization ensemble $\Delta M(h)$ via

$$\sigma^2(h) = \sigma_1^2(h) + N_1(h) \, N_2(h) \, \Delta M^2(h) \qquad (3)$$

where the relative contribution $N_1$, $N_2$ of the two phases can be infered from the mean value measurement

$$\langle M \rangle(h) = N_1(h) \, M_1(h) + N_2(h) \, M_2(h) \qquad (4)$$

using the normalization condition $N_1 + N_2 = 1$.
(The corresponding field variation of the backbending is $2h_m$).
In particular at zero field we have

$$\sigma^2(h_+ = 0) = \sigma_1^2(h_+ = 0) + \Delta M_0^2 \qquad (5)$$

where $\Delta M_0 / N$ is an estimation from the system of N spins, of the magnetization jump at the thermodynamic limit.

## Remarks :
- For the experiment to be feasible the material should allow metastable states, i.e. have an hysteresis cycle.
- The amplitude of the backbending region $2h_m$ is proportional to the surface of the cluster, i.e. the cluster size should be as small as possible.
- Eqs.(1),(2) are exact while eqs.(3),(4),(5) come from a double saddle point approximation (see appendix) that may be not precise for very small lattices. We can do Monte Carlo simulations to check the precision of these expression as a function of the cluster dimension and size .

## Appendix : magnetization fluctuations

The magnetization distribution at a temperature $\beta$ with an external field h reads

$$P_{\beta h}(M) = Z_\beta(M) \exp(\beta M h) / Z_{\beta h} \qquad (6)$$

where $Z_{\beta h}$ is the partition sum of the system in an external field and $Z_\beta(M)$ is the partition sum of the system with fixed magnetization. Using the equation of state

$$<M>(h) = \beta^{-1} \, \partial \ln Z_{\beta h} / \partial h$$

we immediately get the exact relation between the magnetization fluctuation and the magnetic susceptibility for the system in an external magnetic field

$$\chi(h) = \partial <M> / \partial h = \beta \, \sigma^2(h)$$

The constant magnetization ensemble (described by $Z_\beta(M)$) is defined as the ensemble such that all events have exactly the same magnetization.
For this ensemble we can define an effective field

$$h'(M) = -\beta^{-1} \, \partial \ln Z_\beta / \partial M$$

and a magnetic susceptibility

$$\chi'(M) = (\partial h' / \partial M)^{-1} = -\beta \, (\partial^2 \ln Z_\beta / \partial M^2)^{-1}$$

The effective field $h'(M)$ represents the physical field that a sample should feel for its most probable response to be M. Indeed if we note $M_0(h)$ the maximum of the distribution (6) we immediately have the equality

$$h'(M_0) = h$$

We now show that the probability distribution (6) allows to compute thermodynamic properties also for the constant magnetization ensemble.
Explicitly the exact relations hold :

$$h'(M) = h - \beta^{-1} \, \partial \ln P_{\beta h} / \partial M$$
$$\chi'(M) = -\beta \, (\partial^2 \ln P_{\beta h} / \partial M^2)^{-1}$$

which means that a concavity anomaly in the magnetization distribution (bimodality) is exactly equivalent to a negative susceptibility in the constant magnetization ensemble.
In the general case we cannot access the whole probability distribution but only its first moments $<M>(h)$, $\sigma^2(h)$. This reduced information is however sufficient to proof the thermodynamic anomaly of the constant magnetization ensemble.
Indeed let us first consider the case in which the distribution (6) has only one maximum $M_0(h)$ (supercritical temperature or non negligible magnetic field). In this case we can perform a saddle point approximation around the maximum

$$\ln P_{\beta h}(M) = \ln Z_\beta(M) + \beta M h - \ln Z_{\beta h} \cong -(M-M_0(h))^2/(2\sigma^2(h)) - \ln (2\pi\sigma^2(h))^{-}$$

which gives

$$\beta \, \sigma^2(h) \cong -\beta \, (\partial^2 \ln Z_\beta / \partial M^2)^{-1} = \chi'(M_0)$$

This shows that for a monomodal distribution the susceptibilities of the two different ensembles are the same. In the coexistence region the distribution is bimodal. In this case we can write the distribution as $P = P_1 + P_2$

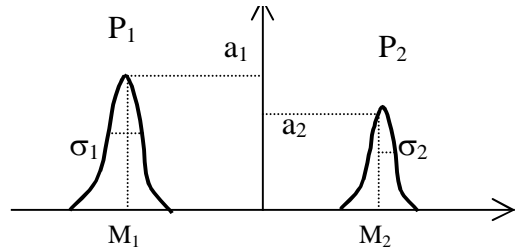

and perform a double saddle point approximation around the two maxima which gives

$$<M>(h) \cong M_1 N_1 + M_2 N_2$$

$$\sigma^2(h) \cong \sigma_1^2 + \Delta M^2 N_1 N_2$$

where we have defined the magnetization jump (i.e. the magnetization interval over which $\chi'$ is negative) $\Delta M = 2(M_1+M_2)$ and we have introduced a normalization $N_i = a_i(2\pi\sigma^2)^-$ i=1,2.
If we neglect the difference between the average and the most probable in the case of single-peaked distributions, these last two equations show that the susceptibility $\chi$ under field is a measure of the susceptibility $\chi'$ for the metastable solution plus the width of the backbending region.
Going towards te thermodynamic limit, the thermodynamic potentials become additive and the distributions $P_i$ converge to an asymptotic expression $p_i$

$$P_i(M) \rightarrow (p_i(m))^N$$

Where m is the magnetization per site. This implies that only the most probable peak survives, and the average values $M_i$ as well as the variances $\sigma_i^2$ scale as the number of spins N

$$<M>/N \rightarrow M_1/N$$

while the total variance goes to zero for all values of the magnetic field but the transition value h=0 ; at zero field the variance converges to the magnetization jump per site

$$\sigma/N \rightarrow \Delta m/2$$